# A Distributed Optimization Approach to the Multi-Regional Day-Ahead Clearing Process in Electricity Markets

Veronica R. Bosquezfoti, Andrew L. Liu, *Member, IEEE*

*Abstract*—The implementation of electricity markets based on locational marginal pricing in a multi-settlement process has allowed wholesale competition, with pricing mechanisms that incentivize the optimal allocation of generation, transmission, and demand-response resources. While efficiency and reliability gains have been achieved in the US at a regional level, the lack of adequate inter-regional coordination mechanisms has limited broader gains. The shortcomings of cross-border coordination of electricity markets become more apparent as the industry transitions towards higher penetration of renewable resources, which tend to be concentrated in certain regions of the country, usually distant from load pockets. In addition to allowing market participants to respond to economic signals beyond the market region where their assets reside, a coordinated market solution would allow the extension of market mechanisms such as financial transmission rights and capacity contracts to cover transactions across borders; thereby, reducing price uncertainty around the investment in new generation resources. This paper presents a distributed solution to the unit commitment problem that allows for full coordination of the market clearing process across interconnected electricity markets.

## I. Introduction

In April 1996, the Federal Energy Regulatory Commission (FERC) issued Order 888 [1], promoting wholesale competition in electric supply through open access to high voltage electric transmission systems. As a result, Independent Transmission Operators (ISOs) were established to administer open access tariffs. Currently, seven ISOs operate in the US, within three largely independent transmission systems (the Eastern, Western and Texas interconnections), serving approximately two thirds of the country's population. In Order 2000 [2], FERC established Regional Transmission Organizations (RTOs) and encouraged utilities to join them [3]. All US ISOs are also established as RTOs, which have operational authority over all transmission facilities under their control. Among the functions of an RTO is management of congestion in the transmission system using market-based mechanisms.

Within each ISO/RTO, the mandated market-based congestion management is implemented using a centrally cleared market based on a series of optimization processes that generate dispatch instructions and price signals to deploy generation and demand-side management resources in the economically optimal manner, while avoiding violation of the physical limitations of transmission and generation facilities. US RTO markets run a two-settlement process that includes day-ahead and real-time markets. The day-ahead market is cleared based on expected conditions, issuing hourly schedules to allow sufficient time for operational planning, and generation start-up and shut-down procedures that may take several hours. The real-time or balancing market adjusts dispatch based on actual system conditions, with updated instructions usually issued with a frequency of five minutes or less.

Where more than one transmission operator exists in an interconnected system, RTOs are also responsible for inter-regional coordination. Considerable amount of electricity is exchanged between interconnected regions in the US. However, the optimization processes utilized to schedule generation and load within RTOs is not extended to the coordination across RTOs. Any existing interregional coordination schemes fall short from approximating the system-wide minimum cost dispatch and are mostly limited to the real-time markets, as described in [4] and [5]. Consequently, generation on/off status is largely determined without optimizing cross-border schedules.

The interregional coordination problem is not unique to the US. There are many other interconnected systems with more than one transmission operator. The various electricity market operators across the European Union, for example, are working towards better interregional coordination with the purpose of optimizing the utilization of transmission interconnections [6]. However, European markets are settled on zonal prices, where a single zone may span an entire national market or a large portion of it, whereas US RTO markets are settled on nodal or bus-level prices.

In 2010, a study performed by ISO New England external market monitor, valued the cost of the inefficiency of transfers across the New York ISO interface at $200 million a year [7]. Considering that the study was performed for a maximum interchange of 500 MW, and that, for example, the hourly average interchange on the MISO-PJM interface was 2.7 GW in 2019 [8], even a small increase in interchange efficiency could bring considerable savings

This paper proposes the use of a distributed optimization algorithm based on the alternating directions method of multipliers (ADMM) in [9]. ADMM is applied to the unit commitment problem in [10], but the solution methods proposed did not reach a feasible solution for our test cases. The algorithm we propose applies an approach developed from the one presented in [11] for the distributed solution of nonconvex optimization problems, taking advantage of the specific structure of the market clearing calculations.

Section II of this paper provides a high-level description of the current interregional coordination practices. Section III presents the formulation of the unit commitment and economic dispatch problems. Section IV presents the consensus optimization distributed solution using ADMM. Section V discusses a heuristic variation of ADMM for nonconvex problems within the context of the proposed application of that algorithm to the solution of the multi-area market clearing process. Finally, Section VI presents simulation results that compare the application of the proposed multi-area solution algorithm to the single-area solution and to a representation of the existing uncoordinated multi-area solution.

## II. CURRENT STATE OF INTERREGIONAL COORDINATION

When more than one RTO operates on an interconnected transmission system, the flow through some transmission elements, especially those close to the boundary, will be driven by generation and load residing in several markets. To avoid overloading those facilities, some coordination must exist among RTOs. The most basic level, illustrated in Fig. 2(a), consists of splitting the transmission capacity of shared facilities. This makes sure that no overloads occur, but may result in sub-utilization of transmission, where one area does not use the entirety of its share while the other could reduce costs by taking over the unused transmission capacity.

In most RTOs, however, coordination occurs past the initial transmission share allocation. Fig. 2(b) illustrates market-to-market procedures, where the capacity of shared constraints is re-allocated between real-time market intervals. More recently, some RTOs have implemented Coordinated Transaction Scheduling (CTS), illustrated in Fig. 1(c), which does not directly re-allocate transmission capacity, but auctions import/export capacity available after the day-ahead market clears. Low participation in CTS has been tied to transaction fees and to the fact that it relies on RTO forecasts of real-time prices at the boundary, which are often inaccurate [8].

None of the existing processes optimize the use of shared transmission capacity in the day-ahead market. As such, decisions regarding which generators will be online are made without taking external transactions into account.

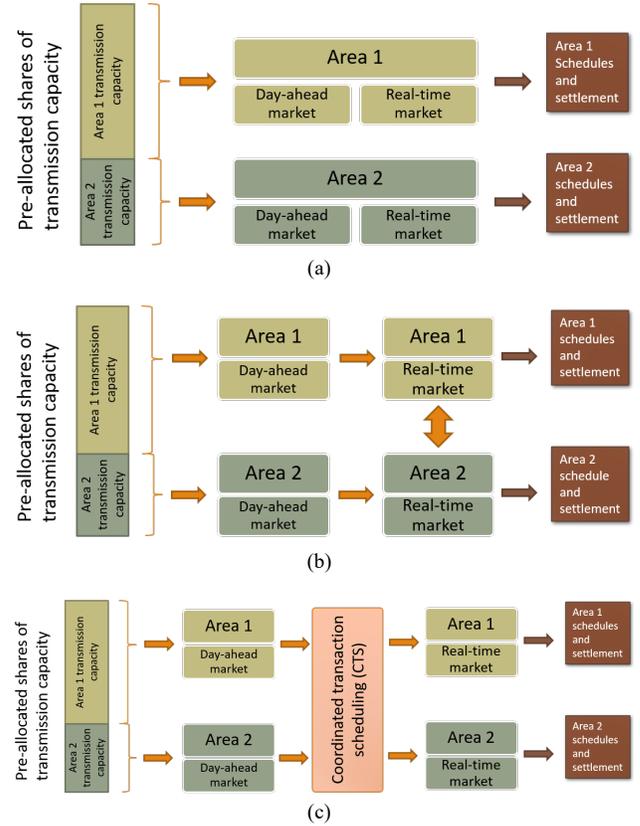

Fig. 1. Existing interregional coordination schemes.

Moving to a single market is unlikely, since RTOs have functions that go well beyond market operations, have a wide range of rules and market processes, and respond to different regulatory entities. Recognizing that, we propose a distributed unit commitment and dispatch approach that allows regions to retain their market rules, clearing and pricing algorithms, and settlement processes.

A multi-regional day-ahead clearing process requires a distributed solution of the unit commitment (UC) problem, which is challenging. However, given the large potential savings, it is worth considering solutions that may require increased computational capabilities. In addition to improving the efficiency of transmission utilization, an effective market-based congestion management across RTO borders would facilitate the investment on new energy sources delivered across regional markets by enabling cross-border capacity contracts and congestion management products. It would also provide price signals that highlight the best candidate locations for investment in new transmission.

## III. THE UNIT COMMITMENT PROBLEM

The two-settlement market clearing process used in US wholesale electricity markets largely consists of UC calculations followed by economic dispatch (ED) calculations. The UC is a mixed-integer program that aims to minimize the total cost of serving load and meeting certain ancillary services requirements while maintaining feasibility of the dispatch with respect to the physical limitations of generation and transmission facilities. The ED calculations optimize the dispatch for a given generation commitment state. Locational Marginal Prices (LMPs) are a byproduct of the ED solution.

In a two-settlement process the RTO calculates and settles day-ahead and real-time markets. The day-ahead market produces forward prices for electricity and provides commitment instructions for generation units that may require advance notice to start up, either because of the physical limitations of the unit, or to plan for adequate fuel reserves and plant staffing. The real-time market adjusts dispatch instructions as needed due to deviations from planned conditions due to forecast errors, unplanned outages, and other unanticipated circumstances.

RTOs typically run UC and ED processes that ensure the feasibility of the dispatch with respect to the limits of generation and transmission facilities under normal operating conditions and under a set of considered contingency scenarios.

The UC problem is an optimal power flow that minimizes the total commitment and dispatch cost.

$$\min f(P,u)$$
Subject to:
$$\begin{aligned} g(P,Q,u,V,\theta) &\leq 0 \\ h(P,Q,u,V,\theta) &= 0 \\ u \cdot P^{min} &\leq P \leq u \cdot P^{max} \\ u \cdot Q^{min} &\leq Q \leq u \cdot Q^{max} \\ u_{g,t} &\in \{0,1\} \quad \forall g,t \end{aligned} \quad (1)$$

Where $P$ and $Q$ are the vectors of real and reactive generator power output, $u$ is the on/off state of each generating unit, and $V \angle \theta$ are the bus voltages. Constraints include nodal power balance equations, flow limits, ancillary service requirements, inter-temporal limits such as minimum run times, minimum down times, and ramp rates.

The ED is similar to the UC calculation, but the generator statuses are fixed, setting the output of all offline units to zero. The output of online units have continuous feasible ranges between each unit's minimum and maximum limits.

$$\min f(P)$$
Subject to:
$$\begin{aligned} f(P,Q,V,\theta) &\leq 0 \\ h(P,Q,V,\theta) &= 0 \\ P^{min} &\leq P \leq P^{max} \\ Q^{min} &\leq Q \leq Q^{max} \end{aligned} \quad (2)$$

The UC model used in this work, shown in Section A of the Appendix, utilizes a DC approximation of the power flow equation and a quadratic generator cost model, resulting in a mixed integer program that can be written as:

$$\min \frac{1}{2} x^T Q x + c^T x + c_u^T u$$
Subject to
$$\begin{aligned} A \begin{bmatrix} x \\ u \end{bmatrix} - b &\leq 0 \\ u &\in \{0,1\} \\ x &\in \mathcal{X} \end{aligned} \quad (3)$$

## IV. Consensus optimization

The alternating directions method of multipliers (ADMM) discussed in [9] is a distributed solution method developed in the 1970s, well suited for convex optimization problems. The consensus optimization problem can be solved using ADMM in a fully parallelizable manner.

Consider the problem:
$$\min \sum_{i=1}^{N} f_i(x_i)$$
Subject to
$$\left. \begin{aligned} x_i - z &= 0 \\ x_i &\in \mathcal{X}_i \end{aligned} \right\} \quad i = 1,...,N \quad (4)$$

The objective function can be decomposed into $N$ parts, which are a function of local variables $x_i$. Local variables are linked by a vector of global variables $z$.

An augmented Lagrangian can be constructed for (4) as:
$$\begin{aligned} L_\rho(x_1,...x_N,z,y) \\ = \sum_{i=1}^{N} \left( f_i(x_i) + y_i^T(x_i - z) + (\rho/2)\|x_i - z\|_2^2 \right) \end{aligned} \quad (5)$$

The updates of the ADMM algorithm are:
$$\begin{aligned} x_i^{k+1} &:= \operatorname*{argmin}_{x_i} \left( f_i(x_i) + y_i^{kT}(x_i - z^k) + (\rho/2)\|x_i - z^k\|_2^2 \right) \\ z^{k+1} &:= \frac{1}{N} \sum_{i=1}^{N} \left( x_i^{k+1} + (1/\rho) y_i^k \right) \\ y_i^{k+1} &:= y_i^k + \rho(x_i^{k+1} - z^{k+1}) \end{aligned} \quad (6)$$

With the average over [1,N] for $x_i^{k+1}$ denoted as $\bar{x}^{k+1}$ (6) can be re-written as:
$$\begin{aligned} x_i^{k+1} &:= \operatorname*{argmin}_{x_i} \left( f_i(x_i) + y_i^{kT}(x_i - \bar{x}^k) + (\rho/2)\|x_i - \bar{x}^k\|_2^2 \right) \\ y_i^{k+1} &:= y_i^k + \rho(x_i^{k+1} - \bar{x}^{k+1}) \end{aligned} \quad (7)$$

The primal ($r$) and dual ($s$) residuals are calculated as:
$$\begin{aligned} r^k &= (x_1^k - \bar{x}^k, ..., x_N^k - \bar{x}^k) \\ s^k &= -\rho(\bar{x}^k - \bar{x}^{k-1}, ..., \bar{x}^k - \bar{x}^{k-1}) \end{aligned} \quad (8)$$

## V. A distributed solution to the unit commitment problem

### A. Solution approach for the distributed mixed integer quadratic program

In [11], a heuristic method is proposed for the solution of nonconvex quadratic programs. The method used in this work is based on that heuristic approach, with a slightly different update rule.

To illustrate the proposed method, assume that, in the problem stated in (4), the feasible regions $\mathcal{X}_i$ are nonconvex.

With each iteration, a convex approximation of (4) is solved, and the solution of each subproblem is projected onto the feasible nonconvex region:

$$\tilde{x}_i^{k+1} := \operatorname*{argmin}_{x_i} \left( f_i(x_i) + y_i^{kT}(x_i - \bar{\tilde{x}}^k) + (\rho/2)\|x_i - \bar{\tilde{x}}^k\|_2^2 \right) \quad (9)$$

$$x_i^{k+1} := P_{\mathcal{X}_i}\left(\tilde{x}_i^{k+1}\right) \quad (10)$$

$$y_i^{k+1} := y_i^k + \rho(\tilde{x}_i^{k+1} - \bar{\tilde{x}}^{k+1}) \quad (11)$$

The updates shown in (9) differ from the updates in [11] in that at each step, the next solution to the convex approximation of the original problem is calculated starting from the previous solution of the convex problem ($\tilde{x}_a^k$) and not from the projection of that solution into the feasible region ($x^k$). Because the consensus ADMM algorithm in (7) is used to solve the convex problem, the algorithm is guaranteed to converge to the optimal solution of the convex approximation.

However, there is no guarantee that the projection of the solution computed from (10) will be found at each iteration. Even if found, there is no guarantee that the corresponding objective value will decrease monotonically. Two of the features of the solution presented in [11] are incorporated into our UC solution to overcome this problem:

i. With each iteration, the feasible solution is compared to the best feasible solution obtained in previous iterations. The best solution is retained.
ii. The algorithm is initialized with a set of randomly generated starting points.

In the proposed algorithm, the projection in (10) consists of an optimization process that ensures that the projected solution is feasible with respect to all constraints in the original problem, including global constraints.

### B. Solution approach for the multi-area unit commitment problem

#### 1) Local decision variables

We solve the UC problem for an interconnected system with several market operators. The set of market operator areas is $\mathcal{A}$.

For the UC problem, the vector of decision variables for each area $a \in \mathcal{A}$ is:

$$x_a = \begin{bmatrix} x_{a,1}^T & \cdots & x_{a,TI}^T \end{bmatrix}^T$$

$$x_{a,t} = \begin{bmatrix} u_t^{(a)} \\ p_t^{(a)} \\ v_t^{(a)} \\ v_t^{H(a)} \\ w_t^{(a)} \\ \theta_t^{(a)} \end{bmatrix} \quad (12)$$

A complete description of the UC formulation used for this work is included in Section A of the Appendix.

In (12), $u_t^{(a)}$ represents the vector of commitment statuses for time period $t$, for all generators in area $a \in \mathcal{A}$. The same notation is used for all other variables that define the hourly status of generators. Similarly, $\theta_t^{(a)}$ represents the phase angles for period $t$ of all buses within, or immediately adjacent to, area $a$.

The nonconvexity of the UC formulation used in work arises from the binary nature of all commitment flags ($u$, $v$, $v^H$, and $w$). Variables $v$, $v^H$, and $w$ are fully determined from the value of $u$; therefore, it is sufficient to enforce $u_{t,g}^{(a)} \in \{0,1\}$ $\forall a,t,g$ to make sure all commitment flags take binary values. To solve the convex approximation of the problem in (9), we relax this constraint to $u_{t,g}^{(a)} \in [0,1]$ $\forall a,t,g$.

#### 2) Global variables

The consensus variables are the phase angles corresponding to the buses included in more than one area solution and the tie-line flows. A tie-line is a branch (transmission line or transformer) where the two ends reside in different areas. Those buses connected to either side of any tie-line are shared buses. We denote the set of shared buses:

$$\mathcal{N}^S = \left\{i : i \in \mathcal{N}^a \cap \mathcal{N}^b\right\} \quad \forall a \neq b, a \in \mathcal{A}, b \in \mathcal{A} \quad (13)$$

The vector of global variables represented by $z$ in (4) is the set of phase angles of all shared buses $\theta^S$ and the set of tie line flows $F^{TL}$.

The vector of shared bus phase angles calculated within the solution of area $a$ is

$$\theta^{S(a)} = \left\{\theta_{i,t}^{(a)} : i \in \mathcal{N}^S\right\} \quad (14)$$

The consensus value of the shared bus phase angles is calculated, for each area $a$, as:

$$\bar{\theta}^{(a)} = \left\{\bar{\theta}_{i,t}^{(a)}\right\}$$

$$\bar{\theta}_{i,t}^{(a)} = \frac{1}{|\mathcal{A}_i|} \sum_{b \in \mathcal{A}_i} \theta_{i,t}^{(b)} \quad \forall i \in \mathcal{N}^S \cap \mathcal{N}^a \quad (15)$$

where

$$\mathcal{A}_i = \left\{a : i \in \mathcal{N}^a\right\}$$

Note that only those shared buses within or neighboring an area $a$ are included in the consensus variable vector for $a$.

The average value of the tie-line flows is calculated as:

$$\bar{F}^{(a)} = \left\{\bar{F}_{ij,t}^{(a)}\right\}$$

$$\bar{F}_{ij,t}^{(a)} = \frac{1}{2}\left(F_{ij,t}^{(a)} + F_{ij,t}^{(b)}\right)$$

$$= \frac{1}{2}\left(B_{ij}\left(\theta_{i,t}^{(a)} - \theta_{j,t}^{(a)}\right) + B_{ij}\left(\theta_{i,t}^{(b)} - \theta_{j,t}^{(b)}\right)\right) \quad (16)$$

$$\forall i,j \in \mathcal{N}^a \cap \mathcal{N}^b, \forall a,b \in \mathcal{A}$$

#### 3) Consensus unit commitment

The update of the convex approximation shown in (9) is calculated for the particular case of the UC problem as:

$$x_a^{k+1} := \underset{x_a}{\operatorname{argmin}} \begin{pmatrix} \frac{1}{2}x_a^T Q_a x_a + c_a x_a \\ +\lambda_a^{kT}(\theta^{S(a)} - \bar{\theta}^{(a)k}) \\ +(\rho/2)\left\|\theta^{S(a)} - \bar{\theta}^{(a)k}\right\|_2^2 \\ +\mu_a^{kT}(F^{TL(a)} - \bar{F}^{(a)k}) \\ +(\rho/2)\left\|F^{TL(a)} - \bar{F}^{(a)k}\right\|_2^2 \end{pmatrix} \quad (17)$$

At each iteration, the residual is calculated as:

$$r_{(a)}^k := \begin{bmatrix} \theta^{S(a)} - \bar{\theta}^{(a)k} \\ F^{TL(a)} - \bar{F}^{(a)k} \end{bmatrix} \quad (18)$$

The dual update in (11) is calculated for the relaxed unit commitment as:

$$\lambda_a^{k+1} := \lambda_a^k + \rho\left(\theta^{S(a)} - \bar{\theta}^{(a)k}\right)$$
$$\mu_a^{k+1} := \mu_a^k + \rho\left(F^{TL(a)} - \bar{F}^{(a)k}\right) \quad (19)$$

The projection of $x_a^{k+1}$ onto the feasible region shown for the general case in (10) is calculated by selecting a commitment threshold $0 < \xi < 1$ and setting the values of the commitment flag $u$ for all generators, periods and areas accordingly. The remaining commitment variables are set based on the hourly commitment.

From the resulting fixed set of commitment variables, an ED calculation is performed to compute the optimal dispatch level for each online generator. The ED is computed for the multi-area case using a consensus algorithm. The ED is a convex problem, so the consensus solution of the ED should converge towards the optimal solution for the fixed commitment.

$$x_a^{ED,k+1} := \underset{x_a^{ED} \in \mathcal{X}_a^{ED}}{\operatorname{argmin}} \begin{pmatrix} \frac{1}{2} x_a^{ED\,T} Q_a^{ED} x_a^{ED} + c_a^{ED} x_a^{ED} \\ + \lambda_{ED\,a}^{kT} (\theta_{ED}^{S(a)} - \bar{\theta}_{ED}^{(a)k}) \\ + (\rho_{ED}/2) \left\| \theta_{ED}^{S(a)} - \bar{\theta}_{ED}^{(a)k} \right\|_2^2 \\ + \mu_{ED\,a}^{kT} (F_{ED}^{TL(a)} - \bar{F}_{ED}^{(a)k}) \\ + (\rho/2) \left\| F_{ED}^{TL(a)} - \bar{F}_{ED}^{(a)k} \right\|_2^2 \end{pmatrix} \quad (20)$$

The economic dispatch update in (20) is similar to (17), but the commitment variables are fixed. The decision variables in the ED problem are:

$$x_a^{ED} = \begin{bmatrix} x_{a,1}^{ED\,T} & \cdots & x_{a,TI}^{ED\,T} \end{bmatrix}^T$$
$$x_{a,t}^{ED\,T} = \begin{bmatrix} p_{ED\,t}^{(a)} \\ \theta_{ED\,t}^{(a)} \end{bmatrix} \quad (21)$$

The multi-area ED is solved using a consensus ADMM algorithm. The consensus value of global variables is calculated as shown in (15) and (16). Residual and dual updates are calculated as shown in (18) and (19).

*4) Heuristic multi-area unit commitment algorithm*

(0) Set $x_{best} = 0$, $C_{best} = \infty$
(1) For $n = 1, \ldots, N_{IC}$
   (a) Randomly generate initial conditions for $x_a^0$, $\lambda_a^0$ and $\mu_a^0$ for all $a \in \mathcal{A}$.
   (b) Calculate consensus values for $\bar{\theta}^{(a)0}$ and $\bar{F}^{(a)0}$ for all $a \in \mathcal{A}$ using (15) and (16).
   (c) For UC ADMM iteration $k = 1, \ldots, N_{UC}$
       (i) For each area $a \in \mathcal{A}$, calculate relaxed UC solution from (17).
       (ii) Calculate commitment statuses from the relaxed UC solution.

$$u_{i,t}^{ED\,k} := \begin{cases} 0 & \text{if } u_{i,t}^k < \xi \\ 1 & f\ u_{i,t}^k \geq \xi \end{cases}$$

   (iii) Set all commitment variables based on $u_{i,t}^{ED\,k}$.
   (iv) Calculate ADMM ED solution
       • Iterate over $m = 1, \ldots, N_{ED}$ using (20) and ED versions of (15), (16), (19) as update rules.
       • Calculate residual from (18). Stop multi-area ED solution if $\left|r_{(a)}^m\right|_\infty \leq \varepsilon_{ED}$.
   (d) Set $x_{best} = x_a^{ED}$ and $C_{best} = C_a^{ED}$ if the new solution is feasible and the cost is lower than the current $C_{best}$.
(2) Use the UC solution corresponding to $x_{best}$ to calculate LMPs via ADMM. Share boundary bus LMPs across areas.

## VI. SIMULATION CASES AND RESULTS

### A. Test scenarios

Three separate scenarios were evaluated for each transmission system model:

*1) Single-area model*

This represents the UC and ED results of the entire system modeled as a single area. This model yields the least-cost feasible solution but requires a single market operator to receive bids and offers from all participants and to maintain a model of the entire interconnected transmission system.

*2) Uncoordinated multi-area model*

The UC and ED results are computed separately for each area, with a price-insensitive interchange modeled across pre-defined interfaces. The interfaces are defined as tie-line flow aggregates with fixed weights. The fixed interface definition weighting factors and the interchange amounts were chosen based on the exchange in the single-area model. The interchange amounts are either fixed for the 24-hour period or have peak and off-peak values.

This model is intended to represent the current coordination methods. While price-sensitive bids and offers may be placed at interchange locations, in practice, the lack of coordination between markets does not guarantee that a transaction to the border in one area will have a corresponding transaction from the border in the other area. Cross border transactions are therefore usually price insensitive. Inter-regional coordination schemes exist in some markets but they are limited to the real-time markets and do not impact day-ahead commitment.

The state of inter-regional coordination and its limitations within the Midcontinent ISO markets is described in the State of the Markets report [8]. The report indicates that external transactions are overwhelmingly scheduled in a non-price sensitive manner, based on the participant expectation of price differences between markets. Consequently, inter-regional transactions can be uneconomic. Also, it indicates that while some participation in the Coordinated Transaction Scheduling (CTS) process has been observed, it is still a very small fraction of cross-border schedules due mostly to inaccuracy on the interface price forecasts generated by the ISOs that are the basis for the CTS process. As such, the most accurate representation of the current state of day-ahead cross-border schedules is a fixed hourly import or export at each interface, with interfaces defined as fixed aggregates of a number of boundary locations.

*3) Coordinated multi-area model*

The UC and ED results are computed using the distributed algorithm presented in Section V. Market operators must clear markets in a coordinated fashion but are only required to share a subset of boundary conditions between iterations. Each area can keep bid and offer information private, may have different market rules, and is only required to maintain an accurate model of the internal transmission system and tie-lines.

*B. Test cases*

Three transmission network models were used to evaluate the proposed algorithm:
1. A 14-bus, two-area case
2. A 200-bus, three-area case
3. A 500-bus, three-area case

All cases are based on transmission network models obtained from the Texas A&M University electric grid test case repository [12]. All generation cost parameters and inter-temporal constraints were added to the 14-bus case. The 200 and 500 -bus base cases included incremental costs, but no commitment costs and parameters. Commitment costs and constraints were computed for the larger cases from information in [13].

*C. System information*

Test cases were executed in Matlab R2020b Update 1, calling IBM ILOG CPLEX Optimization Studio version 12.9. They were run on a PC with an Intel Core i7 6560U CPU@2.2GHz, 16 GB RAM and a 64 bit OS.

*D. Simulation results*

Table I shows the 24-hour commitment cost resulting from each scenario and transmission network model.

TABLE I
TEST CASE RESULTS

| Case | Solution method | Total 24-hour cost (commitment and dispatch) | Execution time (s) |
|---|---|---|---|
| 14-bus | Single-area | $ 161,302 | 0.43 |
| | Uncoordinated | $ 169,269 | 0.30 |
| | Heuristic multi-area | $ 163,770 | 120 |
| 200-bus | Single-area | $ 454,908 | 310 |
| | Uncoordinated | $ 699,935 | 52 |
| | Heuristic multi-area | $ 498,296 | $3.8 \times 10^3$ |
| 500-bus | Single-area | $ 1,168,104 | $3.0 \times 10^3$ |
| | Uncoordinated | $ 1,293,572 | 740 |
| | Heuristic multi-area | $ 1,221,181 | $15 \times 10^3$ |

Starting from the uncoordinated day-ahead clearing model that represents the current practices, the maximum savings are achieved by clearing the entire system as a single area. For the 14-bus case, the proposed algorithm resulted in a cost reduction of 69% of the maximum savings. For the 200 and 500 bus cases, the corresponding cost reductions were 82% and 58%, respectively.

Execution times are noted, highlighting the increased complexity of the proposed method. While there is room for improvement on the implementation of the proposed algorithm in terms of performance, it is worth noting that increased interregional coordination would certainly require additional computational capabilities.

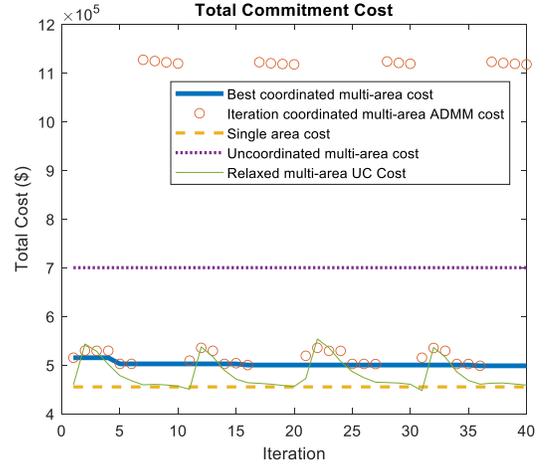

Fig. 2. Total commitment and dispatch cost comparison for the 200-bus cases. The total cost on each iteration is shown in comparison with the single-area commitment cost and the multi-area uncoordinated cost.

The cost results on each iteration for the 200-bus case is shown in Fig. 2. The dotted line is the uncoordinated multi-area cost that represents the current market clearing practices, and the dashed line is the single-area cost that should achieve the lowest cost.

The circle markers show the UC cost calculated on each ADMM iteration. This cost includes the start-up, shut-down and no-load costs calculated from the projected commitment variable values, plus the ADMM dispatch cost in (20). The iteration cost replaces the best cost if it is lower and the ED is feasible. $N_{IC} = 4$, so initial conditions are randomly generated four times. For each randomly generated set of initial conditions, $N_{UC} = 10$ ADMM iterations were completed for the 14 and 200 -bus cases, while for the 500-bus case, $N_{UC} = 12$.

In all simulation results, the best multi-area commitment cost was achieved before the last ADMM iteration. The experience from test cases indicates that the least-cost relaxed UC solution did not result in the least-cost feasible multi-area UC solution. This illustrates why simply relaxing binary variables and applying ADMM to the resulting relaxed UC problem and then projecting the solution onto the feasible region did not produce satisfactory results.

The very high ADMM costs shown in Fig. 2 are associated with UC solutions for which we could not find a feasible multi-area dispatch where boundary conditions across areas match.

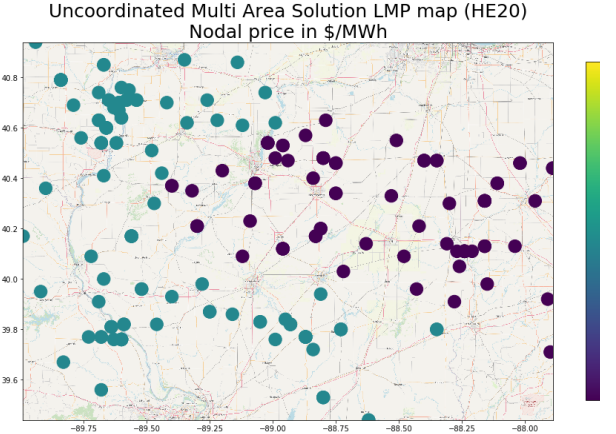

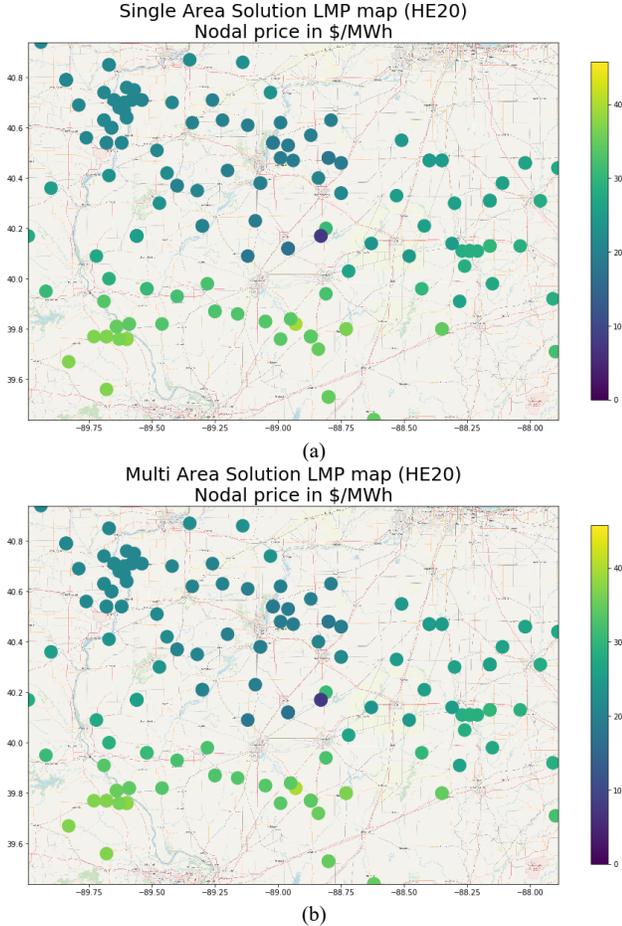

Fig. 3. Hour-ending 20 LMP map comparison for the 200-bus case between (a) the single-area solution and (b) the multi-area coordinated solution.

Fig. 4. Hour-ending 20 LMP map for the uncoordinated multi-area 200-bus case.

Fig. 3 compares the LMP results from the single-area solution and the multi-area coordinated solution applied to 200-bus case for hour-ending 20. While the heuristic multi-area solution did not exactly replicate the single-area results, LMPs do appear to capture the same congestion patterns, creating appropriate price incentives for generators and price-responsive demand to operate in a way that nearly maximizes overall system efficiency.

Fig. 4 shows LMPs for the same period. The fixed transfer in the uncoordinated case results in a sub-utilization of the transmission system capacity. Each area has a single, flat LMP. There is a small difference between areas 1 and 2, while prices in area 3 are near zero. This happens because the entirety of area 3 demand on that period is met by renewable resources. It is worth noting that because the UC minimizes costs including start-up, shut-down, no-load, and dispatch costs over the 24-hour study period, and the LMPs as computed are driven only by the marginal dispatch cost of generation for the current hour, lower LMPs do not necessarily indicate lower total costs.

## VII. Conclusion

In this paper we have demonstrated the application of a heuristic algorithm to the distributed solution of the unit commitment and economic dispatch problems. The algorithm was tested on three cases of varying size, representing transmission systems that contain separate regional markets. The results were compared to the corresponding ideal single-area case and an uncoordinated multi-area case with fixed hourly interchange across markets that represents the current state of interregional coordination. In all cases the proposed algorithm resulted in considerable efficiency gains with respect to the uncoordinated solution, in addition to LMPs across the entire interconnected system that result in an extension of market-based congestion management to cross-border transactions.

The algorithm provides considerable improvement over direct application of ADMM to the relaxed, partially relaxed, or binary unit commitment algorithms found in prior work that failed to find a feasible dispatch. Gains are achieved by capturing the relaxed solution that leads to a feasible solution of the nonconvex problem by (a) iteratively re-initializing the ADMM solution, (b) finding, when possible, a projection of the relaxed solution onto the feasible region with each unit commitment ADMM iteration, (c) verifying that the projected binary solution does not lead to violation of generation and flow constraints or global constraints by running a multi-area economic dispatch calculation with each projected unit commitment solution, and, finally, by (d) retaining the best solution, which typically was not achieved at the optimum of the relaxed problem.

The proposed algorithm could serve as a basis to improve inter-regional coordination, which in addition to increase the efficiency of the use of existing generation and transmission resources, would allow for the inclusion of cross-border transactions in markets for capacity and congestion products.

## Appendix

### A. Unit commitment formulation

The formulation of the UC problem used in this work is adapted from [14] and [15]. Reserves requirements and contingency conditions are not explicitly included in this formulation, but their inclusion is a straightforward extension of the formulation below. A security constrained unit commitment formulation that includes ancillary services can be found in [16].

$$\min \sum_{g,t} \mathcal{C}_g^{UC}(p_{g,t}, u_{g,t}) \quad (22)$$

Subject to:

$$u_{g,t} - u_{g,t-1} = v_{g,t} - w_{g,t} \quad \forall t,g \quad (23)$$

$$v_{g,t}^H \leq \sum_{\tau=t-1}^{\max(1, T_g^{cold} - T_g^D, 0)} w_{g,t} \quad (24)$$

$$\forall t \in [\max(1, T_g^{cold} + T_g^0), TI], g$$

$$v_{g,t}^H \leq v_{g,t} \quad \forall t,g \quad (25)$$

$$u_{g,t} = 1 \quad \forall t \in [1, T_g^U - T_g^0], g : u_g^0 = 1 \quad (26)$$

$$\sum_{\tau=t-T_{min}^U+1}^{t} v_{\tau,g} \leq u_{g,t} \quad \forall t \in [T_g^U, TI], g \quad (27)$$

$$u_{g,t} = 0 \quad \forall t \in [1, T_g^D + T_g^0], g : u_g^0 = 0 \quad (28)$$

$$\sum_{\tau=t-T_{min}^D+1}^{t} w_{\tau,g} \leq 1 - u_{g,t} \quad t \in [T_g^D, TI], g \quad (29)$$

$$\left.\begin{array}{l} p_{g,t} \leq \left(P_g^{max} - P_g^{min}\right) u_{g,t} - \left(P_g^{max} - P_g^{SU.max}\right) v_{g,t} \\ p_{g,t} \leq \left(P_g^{max} - P_g^{min}\right) u_{g,t} - \left(P_g^{max} - P_g^{SD.max}\right) w_{g,t-1} \end{array}\right\} \forall t, g : T_g^U = 1$$

$$\left.\begin{array}{l} p_{g,t} \leq \left(P_g^{max} - P_g^{min}\right) u_{g,t} - \left(P_g^{max} - P_g^{SU.max}\right) v_{g,t} \\ \quad - \left(P_g^{max} - P_g^{SD.max}\right) w_{g,t-1} \end{array}\right\} \forall t, g : T_g^U \geq 2$$

(30)

$$-RD_g \leq p_{g,t} - p_{g,t-1} \leq RU_g \quad (31)$$

$$\sum_{g \in \mathcal{G}_i} \left(P_g^{min} u_{g,t} + p_{g,t}\right) - \sum_{j \in \mathcal{N}} B_{ij} \theta_{j,t} = D_i \quad \forall i,t \quad (32)$$

$$-F_{ij}^{lim} \leq B_{ij}(\theta_{i,t} - \theta_{j,t}) \leq F_{ij}^{lim} \quad \forall i,j,t \quad (33)$$

$$u_{g,t} \in \{0,1\} \quad (34)$$

Indices
$t \in [1, TI]$    Time intervals
$g \in \mathcal{G}$    Generators
$i \in \mathcal{N}, j \in \mathcal{N}$    Buses

Decision variables
$p_{g,t}$    Power output above the minimum for generator $g$ and time period $t$.
$u_{g,t}$    Commitment status for generator $g$ and time period $t$.
$v_{g,t}$    Startup flag for generator $g$ and time period $t$. $v_{g,t} = 1$ indicates that $t$ is the first online period after turning on.
$v_{g,t}^H$    Hot startup flag for generator $g$ and time period $t$.
$w_{g,t}$    Shut down flag for generator $g$ and time period $t$. $w_{g,t} = 1$ indicates that $t$ is the last offline period after shutting down.
$\theta_{i,t}$    Voltage phase angle at bus $i$.

Generator cost
$C_g^Q$    Quadratic cost coefficient for generator $g$
$C_g^L$    Linear cost coefficient for generator $g$
$C_g^{NL}$    No-load cost for generator $g$ (fixed cost incurred per period when online)
$C_g^{SU}$    Startup cost for generator $g$ (fixed cost incurred on the period when the generator comes online)
$C_g^{HS}$    Startup cost for generator $g$ when it has been offline less than $T_g^{cold}$
$C_g^{SD}$    Shut down cost for generator $g$ (fixed cost incurred on the last online period before the generator shuts down)

Generator parameters
$P_g^{min}$    Minimum output (MW)
$P_g^{max}$    Maximum output (MW)
$T_g^U$    Minimum up time (intervals)
$T_g^D$    Minimum down time (intervals)
$T_g^{cold}$    Cold startup time (intervals)
$RU_g$    Maximum ramp-up rate (MW/interval)
$RD_g$    Maximum ramp-down rate (MW/interval)

Transmission system parameters
$B$    Admittance matrix
$F_{ij}^{lim}$    Flow limit of the branch connecting buses $i$ and $j$.
$D_i$    Demand at bus $i$.

In this formulation of the unit commitment problem, the total power output of each generator during a particular time interval is:

$$P_{g,t} = P_g^{min} u_{g,t} + p_{g,t} \quad (35)$$

The total commitment and dispatch cost in (22) is quadratic function given by:

$$\begin{aligned} \mathcal{C}_g^{UC}&(P_{g,t}, u_{g,t}, v_{g,t}, v_{g,t}^H, w_{g,t}) \\ &= C_g^Q P_{g,t}^2 + C_g^L P_{g,t} + C_g^{NL} u_{g,t} + C_g^{SU} v_{g,t} \\ &\quad + (C_g^{HS} - C_g^{SU}) v_{g,t}^H + C_g^{SD} w_{g,t} \\ &= C_g^Q (p_{g,t} + P_g^{min} u_{g,t})^2 + C_g^L (p_{g,t} + P_g^{min} u_{g,t}) \\ &\quad + C_g^{NL} u_{g,t} + C_g^{SU} v_{g,t} \\ &\quad + (C_g^{HS} - C_g^{SU}) v_{g,t}^H + C_g^{SD} w_{g,t} \\ &= C_g^Q p_{g,t}^2 + (2 C_g^Q P_g^{min} + C_g^L) p_{g,t} \\ &\quad + (C_g^Q (P_g^{min})^2 + C_g^L P_g^{min} + C_g^{NL}) u_{g,t} \\ &\quad + C_g^{SU} v_{g,t} + (C_g^{HS} - C_g^{SU}) v_{g,t}^H + C_g^{SD} w_{g,t} \end{aligned} \quad (36)$$

The constraints in this formulation enforce the physical limits of generation and transmission facilities:
- (23), (24) and (25) set the startup and shut down flags.
- (26) and (27) enforce generation minimum running times.
- (28) and (29) enforce generation minimum down times.
- (30) enforces generator output limits, including startup and shut down limits, and sets the output above minimum $p_{g,t}$ to zero when the generator is offline.
- (31) enforces generator ramp rates.
- (32) represents the power balance equations in a DC power flow model, which is commonly used in US electricity markets [17]. The admittance matrix ($B$) is an $N \times N$ matrix, where $N = |\mathcal{N}|$. The elements of the admittance matrix are calculated as follows:

$$B_{ij} = -\frac{1}{x_{ij}} \quad \forall i \neq j$$

$$B_{ii} = \sum_{\substack{j \in \mathcal{N} \\ j \neq i}} -\frac{1}{x_{ij}}$$

where $x_{ij}$ is the reactance of the branch(es) connecting buses $i$ and $j$.
- (33) enforces branch flow limits in the positive and negative directions. In this model, the assumption is that only branch limits are monitored, therefore the positive and negative limits are the same and all flow constraints are monitored in both directions. Flowgates and directional flows are a trivial extension of this constraint.

### B. Economic dispatch formulation

The ED formulation used in the simulated market clearing process is a reduced version of the UC formulation that fixes the value of the commitment variables. The dispatch cost for a generator is calculated as:

$$\mathcal{C}_g^{ED}(P_{g,t}) = C_g^Q P_{g,t}^2 + C_g^L P_{g,t} \quad (37)$$

Constraints (30) - (33) are enforced replacing all commitment variables with fixed values. The remaining constraints are not needed for the ED problem.